# Electronic coupling in colloidal quantum dot molecules;

# The case of CdSe/CdS core/shell homodimers

Yossef E. Panfil[1], Doaa Shamalia[1], Jiabin Cui[1], Somnath Koley[1], Uri Banin[1]*

[1] Institute of Chemistry and The Center for Nanoscience and Nanotechnology, The Hebrew University of Jerusalem, Jerusalem 91904, Israel.

* Corresponding author. Email: uri.banin@mail.huji.ac.il (U.B.)

**Abstract**

Coupled colloidal quantum dot molecules composed of two fused CdSe/CdS core/shell sphere monomers were recently presented. Upon fusion, the potential energy landscape is changing into two quantum dots separated by a pre-tuned potential barrier with energetics dictated by the conduction and valence band offsets of the core/shell semiconductors, and width controlled by the shell thickness and the fusion reaction conditions. In close proximity of the two nanocrystals, orbital hybridization occurs, forming bonding and anti-bonding states in analogy to the hydrogen molecule. In this study we examine theoretically the electronic and optical signatures of such a quantum dot dimer compared to its monomer core/shell building-blocks. We examine the effects of different core sizes, barrier widths, different band offsets and neck sizes at the interface of the fused facets, on the system wave-functions and energetics. Due to the higher effective mass of the hole and the large valence band offset, the hole still essentially resides in either of the cores breaking the symmetry of the potential for the electron as well. We found that the dimer signature is well expressed in a red shift of the band gap both in absorption and emission, in slower radiative lifetimes and in an absorption cross section which is significantly enhanced relative to the monomers at energies above the shell absorption onset, while remains essentially at the same level near the band-edge. This study provides essential guidance to pre-design of coupled quantum dot molecules with specific attributes which can be utilized for various new opto-electronic applications.

## Introduction

Colloidal semiconductor nanocrystal Quantum Dots (CQDs) are being explored for over 30 years[1], and a hallmark size dependent property of these systems is the blue shift and discretization of the energy levels in the conduction and valence band edges governed by the quantum confinement effect[2,3]. This is clearly manifested in the narrow features of the CQDs absorption spectrum, created by restricted selection rules for allowed transitions, accompanied by the blue shift of the band gap upon reducing CQD size. Quantum confinement also leads to narrow emission linewidths and emission wavelength that can be tuned by the size of the nanocrystal in analogy to the "particle in a box" problem[4]. While a bare nanocrystal is a poor emitter because of surface traps created by dangling bonds on its surface, a core-shell type-I heterostructure greatly enhances the quantum efficiency by confining the wave-functions of the charge carriers to the core region away from the surface[5–9]. Core-shell CQDs are nowadays bright and stable narrow linewidth nano-emitters reaching nearly 100 percent quantum yield efficiency, and are being used in various opto-electronic applications including in LEDs, LCD displays, lasers, as single photon sources and in photovoltaics[1,10–14]

Over the years, many other morphologies rather than core-shell sphere were synthesized. Among them, dot in rod[15–18], rod in rod[19], core-crown nanoplatelets[20], quantum well nanoplatelets[21–24] and nano-dumbbells[25–27]. These morphologies served as a testbed for analyzing the effects of dimensionality on the electronic and optical properties of nanostructures. While such morphologies change the dimensionality of the nanocrystal, none of the above is testing the interaction between two CQDs at short distance. In addition, the prior complex colloidal semiconductor nanostructures that addressed coupling effects are so far based on coupling through organic linker which poses a high potential barrier[28–30], or either on epitaxial growth on tips of a rod or on an existing facet of a seed crystal [31]which is limiting the complexity and the tunability of coupled colloidal quantum dot systems.

Recent work by Cui *et al.* established a method to couple two CQDs chemically thus forming electronically coupled CQD molecules (CQDM)[32] . Briefly, the approach starts from two core-shell CdSe/CdS building blocks, which are initially bound by a linker molecule. The two core/shell CQDs are then being fused at higher temperatures. Figure 1a presents a transmission electron microscopy (TEM) image of such a homodimer CQD Molecule made out of two core-shell CdSe/CdS CQDs. The core diameter is 2.8nm and the shell thickness is 2.1nm. Fig. 1b shows a high-resolution image in high angle annular dark field-scanning TEM (HAADF-STEM) mode in which the continuity of the atomic planes of the different CQDs are clearly resolved confirming the fusion process. Commensurate energy dispersive X-Ray spectroscopy (EDS) mapping of the different elements confirms the presence of the two CdSe cores separated by twice the shell thickness (Fig. 1c-f).

As CQDs are often called "artificial atoms", two attached CQDs can form an artificial molecule. Such a homodimer CQDM is thus analogous to the hydrogen diatomic molecule in which, at a short distance between the hydrogen atoms, bonding and anti-bonding states form by the hybridization of the 1S states. While studies about the coupling of two dots grown by molecular beam epitaxy (MBE) methods have been done[33–35], almost no such studies are reported for CQDMs. Importantly, the CQDMs are characterized by smaller size and enhanced quantum confinement effects relative to the MBE dots. This, combined with the ability to engineer the distance between the two CQDs, results in hybridization energies that well compete with $K_bT$ even at room temperature, while in the structures grown by MBE the coupling is usually limited to cryogenic temperatures. In addition, the above described method can be used also to create fine-tuned complex potential landscapes mixing type I and Type II heterostructures within a CQDM offering a path for a variety of coupled systems. Additionally, CQDMs offer flexibility in integration within various solvents, polymer films, and printing and patterning approaches, via control of their surface chemistry. All together, these CQDMs open the way for various new applications utilizing the electronic coupling between two emission centers. For example, dual-color emitters, electric field sensors and quantum gates for quantum computation applications.

Upon fusion, the potential energy landscape is changing from the core/shell type I heterostructure in which a core composed of the smaller band gap material (CdSe) is embedded inside the shell of the larger band gap material (CdS), to two closely spaced quantum dots separated by the barrier with height dictated by the band offsets between CdSe and CdS (Fig. 1.g). The conduction band (CB) offsets between CdSe to CdS is still under debate and ranges between 0.32eV to -0.1eV[36-40]. In the case of a shallow conduction band offset, for example 0.1eV, and for core diameter below 4nm, the confinement energy for the first electron level in the conduction band is higher than 0.1eV leading to a "quasi-type II" situation. Focusing on this condition as a first example - the hybridization of the lowest energy conduction band wave-functions of the monomers becomes accessible. The resulting wave-functions are those of a symmetric state (red) and anti-symmetric state with a node in the center (green). Commensurate to the conduction band variations, the valence band (VB) offset in this system ranges between 0.42eV to 0.74eV. In addition, the hole effective mass is much larger than that of the electron. As a result, the picture for the hole is still that of essentially two separate CQDs which means that in the single exciton regime the hole wave-function is mostly localized inside one of the cores (blue). Upon taking into account the electron-hole coulomb interaction, the first electron wave-function becomes more localized in the core where the hole resides, while the next electron level will be more localized in the other core (red and green dashed lines in Fig. 1g).

One key observation of Cui *et al.* was a red shift of the emission wavelength upon fusion indicative of quantum coupling. As in the well-known tight binding example of symmetric double square quantum well, the CB $1S_e$ ground level of each CQD is shifted to lower energies due to the presence of the other CQD[41]. However, this shift does not lift the twofold degeneracy of the lowest CB state. This degeneracy is lifted only by the coupling between the two CQDs ($\Delta E_{1-2}^d$ in Fig. 1h). The fusion energy, which is the red shift between the ground states before and after fusion, is marked here as $\Delta E_f$. Taking into account the Coulomb energy due to the other charge carrier and the charges on the surface due to dielectric mismatch between the nanocrystal and the surroundings (magenta levels in Fig. 1h), the monomer ground state energy is red shifted by $C_m$ that is greater than $C_d$, the Coulomb red shift of the dimer ground state, because of the stronger delocalization of the electron in the dimer case. The difference between the Coulombic terms $\Delta E_c$ defined as:

$$\Delta E_c = C_m - C_d. \quad (1)$$

The total red shift of the band gap energy of the dimer with respect to the monomer is then:

$$Red\ shift = \Delta E_f - \Delta E_c. \quad (2)$$

In this study, we examine the electronic and optical signatures of homodimer CQDMs with respect to their CdSe/CdS core/shell monomer building blocks using a theoretical approach. We first study how the different sizes of the core, shell barrier height and width, and the neck thickness, which is formed at the interface of the CQDs, affect the coupling energies and the wave-functions. We then take into account the Coulomb energy to study the emission red shift and compare it with the experimental observations. We also calculate the excited states in a CQDM dimer and show that the fusion can further be identified by the changes in the absorption spectrum.

## Methods

We have calculated the energy levels and the wave-functions using the single band effective mass approximation and the self-consistent Schrodinger-Poisson equations solved on a mesh by COMSOL Multiphysics[36,42,43](See supplementary materials for the mesh details). Briefly, we start by building the 3D geometry of the CQD or the CQDM. The geometry contains different domains: for example, the core and the shell. For each domain we define the following material parameters. The electron and hole effective mass $m_e^*$ and $m_h^*$, the conduction and valence band potential $V_c$ and $V_v$ and the relative dielectric constant $\varepsilon$ (see Table I). For the valence band we used anisotropic effective masses corresponding to the Wurtzite structure of the semiconductors[32]. The influence of strain between the CdSe/CdS core/shell is the reduction the CB offset while increasing the VB offset[44]. This effect is not taken in our calculation explicitly, but can be inferred by the different band offsets. The fusion direction in our simulation is along the c-axis of the monomers. We start by solving the Schrodinger equation for the electron and hole neglecting any coulomb interaction:

$$\left(-\frac{\hbar^2}{2}\nabla\left(\frac{1}{m_e^*(r)}\nabla\right)+V_c(r)\right)\Psi_{e_n}^{m/d}(r)=E_{e_n}^{m/d}\Psi_{e_n}^{m/d}(r). \qquad (3)$$

$$\left(-\frac{\hbar^2}{2}\nabla\left(\frac{1}{m_h^*(r)}\nabla\right)-V_v(r)\right)\Psi_{h_n}^{m/d}(r)=E_{h_n}^{m/d}\Psi_{h_n}^{m/d}(r). \qquad (4)$$

Where $E_{e_n}^{m/d}$ and $\Psi_{e_n}^{m/d}$ are the n$^{th}$ energy eigenvalue and the wave-function in the conduction band of the monomer/dimer, respectively, neglecting any coulomb interaction (same for the VB energies, $E_{h_n}^{m/d}$ and wave-functions, $\Psi_{h_n}^{m/d}$). These equations are solved using the BenDaniel-Duke boundary conditions at the interfaces. The entire computational space is extended away from the CQD outer surface by defining a box containing the CQD in its center. The Dirichlet boundary condition is then applied to the surfaces of the box.

In order to account for the coulomb energy due to the other charge carrier, we solve the Poisson equations:

$$\nabla\left(\varepsilon_0\cdot\varepsilon(r)\nabla\phi_e(r)\right)=-q<\Psi_{e_1}^{m/d}(r)|\Psi_{e_1}^{m/d}(r)>. \qquad (5)$$

$$\nabla\left(\varepsilon_0\cdot\varepsilon(r)\nabla\phi_h(r)\right)=q<\Psi_{h_1}^{m/d}(r)|\Psi_{h_1}^{m/d}(r)>. \qquad (6)$$

Where $\phi_{e/h}$ are the electric potentials accounting for the electron-hole interaction. In addition to the potential that each charge carrier sees in the presence of the other charge carrier, it also sees its own self polarization $\phi_{e/h}^{sp}$ due to the dielectric mismatch with the surroundings. This potential is calculated by[43]:

$$\phi_{e/h}^{sp} = \phi_{e/h} - \phi_{e/h}^{H}. \qquad (7)$$

Where $\phi_{e/h}^{H}$ is the potential produced by the charge carrier as if it was in a homogenous dielectric environment of the nanocrystal.

$$\nabla\big(\varepsilon_0 \cdot \varepsilon_{CdS}\nabla\phi_e^{H}(r)\big) = -q < \Psi_{e_1}^{m/d}(r)|\Psi_{e_1}^{m/d}(r) >. \qquad (8)$$

$$\nabla\big(\varepsilon_0 \cdot \varepsilon_{CdS}\nabla\phi_h^{H}(r)\big) = q < \Psi_{h_1}^{m/d}(r)|\Psi_{h_1}^{m/d}(r) >. \qquad (9)$$

Then, the above potentials are introduced into the Schrodinger equation and solved in a self-consistent manner until the eigen-energies $E'_{e/h_n}{}^{m/d}$ converge.

$$\left(-\frac{\hbar^2}{2}\nabla\left(\frac{1}{m_e^*(r)}\nabla\right) + V_c(r) + q(\phi_h + \phi_e^{sp})\right)\Psi_{e_n}'{}^{m/d}(r) = E'_{e_n}{}^{m/d}\Psi_{e_n}'{}^{m/d}(r). \qquad (10)$$

$$\left(-\frac{\hbar^2}{2}\nabla\left(\frac{1}{m_h^*(r)}\nabla\right) - V_v(r) - q(\phi_e + \phi_h^{sp})\right)\Psi_{h_n}'{}^{m/d}(r) = E'_{h_n}{}^{m/d}\Psi_{h_n}'{}^{m/d}(r). \qquad (11)$$

The Coulomb energies $C_m$ and $C_d$ are then calculated by:

$$C_{m/d} = \frac{|E'_{e_1}{}^{m/d} - E_{e_1}{}^{m/d}| + |E'_{h_1}{}^{m/d} - E_{h_1}{}^{m/d}|}{2}. \qquad (12)$$

And $\Delta E_f$ is calculated by:

$$\Delta E_f = E_{e_1}{}^{m} - E_{h_1}{}^{m} - (E_{e_1}{}^{d} - E_{h_1}{}^{d}). \qquad (13)$$

The overlap integral between an electron in state $E'_{e_n}{}^{m/d}$ to hole in state $E'_{h_n}{}^{m/d}$ is:

$$overlap\ integral = < \Psi_{h_n}'{}^{m/d}|\Psi_{e_n}'{}^{m/d} >^2. \qquad (14)$$

**Table I.** Material parameters used in the simulations.

|  | CdSe | CdS | Environment | Units | Ref. |
|---|---|---|---|---|---|
| $V_c$ | 0 | 0-0.32 | 5 | [eV] | 36,38–40,45 |
| $V_v$ | 0 | -0.42-(-0.74) | -5 | [eV] | 36,38–40,45 |
| $m_e^*$ | 0.112 | 0.21 | 1 | $m_0$ | [46] |
| $m_{h\perp}^*$ | 0.48 | 0.376 | 1 | $m_0$ | [46] |
| $m_{h\parallel}^*$ | 1.19 | 0.746 | 1 | $m_0$ | [46] |
| $\varepsilon_\perp$ | 9.29 | 8.28 | 1 | - | [46] |
| $\varepsilon_\parallel$ | 10.16 | 8.73 | 1 | - | [46] |

**Results and discussion**

We start by analyzing the coupling energy ($\Delta E_{1-2}^d$ in Fig. 1h) dependence on different diameters of CdSe cores and different barrier widths. The barrier width is controlled by overlapping the two outer spheres of the core/shell CQDs (Fig. 2d). In order to emphasize solely the effects of the barrier width and core diameters, we maintained the overall CQD monomer diameter size to be 17nm, large enough such that even in the largest barrier width and the largest core size (where center to center distance is maximized) the resultant neck width will not have significant influence (Fig. 2a-c). The neck width effect will be discussed later on. As mentioned before, the literature value of the conduction band offset between CdSe to CdS is varying between 0.32eV to -0.1eV. Moreover, in previous theoretical calculations, strain was found to reduce the CB offset via the deformation potential[44].Thus, we examined three representative CB offset values of 0.32eV, 0.1eV and 0eV.

For the case of 0eV conduction band offset, because of the higher effective mass of CdS compared to CdSe, the electron wave-function is more concentrated in the shell (Fig. 2a). For 0.1eV and 0.32eV conduction band offset energies, the electron wave-function is concentrated around the core but delocalized also into the shell (Fig. 2b-c). As expected, the general trend is that as the core diameter and the barrier width decrease, the coupling energy $\Delta E_{1-2}^d$ increases. In addition, as the band offset decreases the coupling energy increases (points E1, F1 and G1 in Fig.2e-g, $\Delta E_{1-2}^d$ varies between 1-9meV). However, in small core diameters and small barrier widths the trend is opposite. As the band offset becomes higher the coupling energy increases (points E2, F2 and G2 in Fig.2e-g, $\Delta E_{1-2}^d$ varies between 16-35meV).

In order to understand this behavior, we plot the potential energy landscape together with the eigen-energies for the above-mentioned points (Fig.2 E1-G2). Only for the case of 0.32eV band offset the eigen-energies are below the band offset. In large cores and thick barriers (points E1, F1 and G1), there is an effective high barrier due to the confinement energies and large distances. This limits tunneling-coupling and hybridization is small. However, in small cores and thin barrier width (points E2, F2 and G2), with low conduction band offsets the box is essentially an elliptical rod like architecture rather than two coupled dots and the spacing between the two low lying states is rather small (points E2, F2). For a larger band offset (point G2), in the case where the eigen-energies are below the band offset, the effective confinement box becomes smaller (the cores themselves) and hence leads to a bigger separation between the symmetric and anti-symmetric hybridized states. However, even in this case, when the core size is too small, for example, 2nm core and ~1nm barrier thickness, the coupling energy decreases again because the electron eigen-states are above the band offset.

We next examined how the neck size affects the coupling energy. For this we use a structure of two CQDs with core diameter of 2.8nm and shell thickness of 2.1nm creating an overall CQD with 7nm diameter. In order to control the neck, we are attaching two CQDs at a center to center distance of 7nm so their surfaces touch. Then we converted half of the spheres in the side which connects the two CQDs, to half ellipsoids so they will overlap each other. We then merge them and hence the neck size is dictated by the long axis of the ellipsoid. As a result, the neck size is varying between 0nm - where the long axis of the ellipsoid is equal to the radius of the sphere (3.5nm), to 7nm - where the long axis of the ellipsoid is infinite.

Starting with a band offset of 0.1eV (blue line and points in Fig. 3a), at a neck size of 7nm, the coupling energy $\Delta E_{1-2}^d$ is 16meV (point B3 in Fig.3a). Reducing the neck size to 4nm reduces the coupling energy by ~70% to only 5meV (point B2 in Fig.3a). Further reduction of the neck to 1.5nm reduces the coupling energy even more, to only 1.5meV (point B1 in Fig.3a). In comparison, for a band offset of 0.32eV (red in Fig. 3a), and neck size of 7nm, because of the higher potential barrier, the coupling energy is only 6meV (point C3 in Fig.3a). Reducing the neck size to 4nm reduces the coupling energy by less than 40% to 3.5meV (point C2 in Fig.3a). Further reduction of the neck to 1.5nm again leads to nearly complete vanishing of the coupling energy (point C1 in Fig.3a).

In order to further comprehend the different behavior for 0.1eV compared to 0.32eV CB band offsets, the wave-function of the symmetric and anti-symmetric states are presented in Fig 3b-c. For 0.1eV with the largest neck, the wave-functions is delocalized all over the shell. Thus, the reduction of the neck size from 7nm to 4nm is significantly reducing the coupling energy. Whereas, for 0.32eV band offset in all neck sizes the wave-function is concentrated around the cores, so the reduction of the neck from 7nm to 4nm is not affecting the coupling energy as dramatically as in the case of 0.1eV band offset. These calculations demonstrate that the neck has a major effect on the coupling energy and filling the neck by suitable fusion reaction conditions can thus change the emission red shift and additional quantum coupling effects significantly.

It is worth noting that the energy scale for the coupling energy $\Delta E_{1-2}^d$ is on the order of tens of meV for the small core and shell sizes leading to thin barriers. In most previous examples of coupled QDs grown by MBE methods the coupling energy that was reported was on the order of less than 1meV[47–49]. The reasons are the relatively larger dot sizes and barrier thicknesses typically attainable for MBE QDs. The coupling energies calculated for CQDM can exceed the room temperature $K_b T$. This allows to identify and utilize coupling effects even without reverting to cryogenic temperatures. These coupling energies can be measured by scanning tunneling spectroscopy where the electron energy levels can be probed without the presence of the hole[33,45,46].

Upon optical excitation of a CQDM, an electron-hole pair (exciton) is generated. As discussed above, the VB offset between CdSe to CdS is larger than that of the CB. In addition, the hole effective mass is larger - so little to no hybridization is expected and calculated for the hole. In this condition the hole is in either of the dots and the potential landscape for the electron is no longer symmetric. In order to study how the coulomb interaction affects the optical properties of the CQDM (dimer) compared to the CQD (monomer), we used 7nm CQDs fused by a neck with width of 5nm where the core diameter is changing between 1nm to 6nm (scheme in Fig. 4d).

The radiative lifetime in CQDs, an additional measurable observable that can indicate on coupling effects, is inversely proportional to the square of the overlap integral between the electron and hole[52]. The overlap integral for the lowest electron state and the highest hole state both for monomer and dimer and for both band offsets was calculated according to Eq. 14 and the results are presented in Fig. 4a. A general trend of non-monotonic overlap integral values holds for all cases. For core sizes up to ~2.5nm the overlap integral decreases with size, while for larger core sizes the overlap integral increases with size. For both band offset values, the overlap integral is slightly smaller in the case of the dimers compared to the monomers. In addition, for all core diameters the overlap integral is higher in the case of 0.32eV band offsets compared to 0.1eV.

In order to understand these trends one should examine the wave-functions of the lowest electron state (equivalent to the lowest unoccupied molecular orbital in molecules, LUMO) and the highest hole state (equivalent to the highest occupied molecular orbital in molecules, HOMO) as a function of core size. A representative set of core sizes is shown: 1nm, 1.9nm and 4.2 nm (A, B and C vertical dashed line in Fig. 4b, the wave functions are presented in Fig.4e-f). Starting with 0.1eV band offset (blue frame in Fig. 1e), in 1nm core diameter the lowest electron wave-function is more localized in the core which hosts the hole and is not symmetric anymore. However, in such a small core also the hole starts to be delocalized. As the core becomes larger, up to 2.5nm core diameter, the hole wave-function is becoming more localized into the core. However, the electron wave-function is still delocalized over the shell (Fig. 4e.B). This leads to a higher overlap in small core diameters compared to larger ones. For core sizes greater than 2.5nm the electron wave-function is becoming more localized in the core region as the core size increases. This leads to a better electron-hole overlap as the core size increases (Fig. 4e.C). Comparing monomer to dimer, one can see that the overlap integral in general is smaller for dimers as expected from the larger volume and the effects of coupling for the CB state and hence more delocalized electron wave-functions, while the hole remains still localized.

The same trend holds also for the 0.32eV band-offset (Fig. 4f). The main difference is that the overlap integral is larger for all core diameters because of the stronger localization of the electron wave-function. In addition, for small core diameters the hole is more delocalized into the shell due to the correspondingly smaller valence band offset (case A in Fig. 4f).

Commensurately, the opposite trend holds for the Coulombic interactions. In small core diameters up to 2.5nm the Coulombic interactions $C_m$ and $C_d$ are increasing with core diameter size. Above 2.5nm, the Coulombic interactions decrease with core size. For all core sizes the Coulombic interactions are stronger in the case of monomer compared to dimer. This is related to the more delocalized electron wave-function in the dimer case reducing the overlap. As for the different band offsets, except from core sizes below 1.5nm, $C_m$ and $C_d$ are larger in the larger band offset of 0.32eV both for monomers and dimers.

By looking on the wave-functions (Fig.4e-f) one can explain the trends. The Coulombic interaction is larger if one or both wave-functions are more localized in space. Starting with 1nm core diameter the hole wave-function is also slightly delocalized in the shell (Fig. 4e-f case

A). By increasing the core diameter, the hole wave-function becomes more localized and hence, $C_m$ and $C_d$ are growing. For core sizes larger than 2.5nm, because of the growing core diameter both the electron and the hole wave-functions spreads on a bigger volume and this fact is decreasing the Coulombic energies.

Comparing the different band offsets, as expected, higher band offset is localizing the charge carriers more and therefore $C_m$ and $C_d$ are bigger. Only at very small core diameters below 1.5nm the values for 0.32eV band offset become smaller. This is because the valence band offset is smaller, and the hole becomes delocalized as well leading to smaller $C_m$ and $C_d$ than for the case of 0.1eV band-offset.

The total emission red shift will be the difference between $\Delta E_f$ , the fusion energy, to $\Delta E_c$, the difference between the Coulombic terms of the monomer $C_m$ compared to the dimer $C_d$. These two quantities are opposite. $\Delta E_f$ will red shift the dimer band gap compared to the monomer. However, since the Coulomb energy is larger for the monomer, $\Delta E_c$ blue shifts the dimer compared to monomer. By looking on $\Delta E_f$ and $\Delta E_c$ as a function of the core diameter, one can see that both for 0.32eV and for 0.1eV band offsets, $\Delta E_f$ is always larger than $\Delta E_c$ (Fig. 4c). As a consequence, one should expect a red shift but never blue shift upon fusion (Fig. 4d). Indeed, experimental emission wavelength is red shifted by the same amount in the case of CQDMs with various sizes compared to their monomer building blocks[32].

We next examine the dependence of $\Delta E_f$ itself on the core diameter. In small core diameters, $\Delta E_f$ decreases while increasing the size because the electron becomes more localized to the core which leads to lesser hybridization. However, beyond 4nm core diameter, the surfaces of the cores get closer leading to more coupling and hence to higher fusion energy. These results also suggest that one of the signatures of fusion will be the emission red shift.

As a point of comparison to the experimental data, we show in Figure 5a the absorption and emission spectra comparing the monomers (core diameter/shell thickness 2.8/2.1nm; blue lines), the unfused dimers (green lines), and the fused dimers (red lines). A clear red shift in the fused dimers is seen in both absorption and emission spectra, accompanied by broadening of the peaks. Firstly, we consider possible contribution of fluorescence resonant energy transfer (FRET) to the shift. This may be expected in a dimer with somewhat different core sizes where the larger core serves as an acceptor, red shifted from the smaller core serving as a donor. Note that the red shift is prominent only in the fused dimers compared to the unfused dimers for which the red shift is nearly identical to the monomers. This absence of observable significant red shift in the emission of the unfused dimers compared to the monomers, indicates little role of energy transfer affecting the spectral shifts, although the energy transfer mechanism may play other roles in the coupling effects within the dimers[32]. Moreover, we find a good agreement between the quantum mechanical calculations and the experimental red shift[32]. In the case of dimers formed from large core/shells (radius of 1.9nm/4nm), no measurable red shift is identified, compared to the case of dimers formed from small core/shells (radius of 1.4nm/2.1nm), where the experimental red shift is ~13meV, both results being in good agreement with our calculations.

An additional signature for fusion is the absorption cross-section. The absorption cross-section, $\sigma$, of the monomers and their corresponding fused and non-fused dimers was extracted from the absorption spectrum and ICP-AES (inductively coupled plasma-Atomic emission spectroscopy) measurements (See supplementary materials for more details)[53–55]. The absorption cross section of non-fused dimers (green Fig. 5a) closely follows the wavelength dependence of the cross section for monomers (blue) and is merely doubled as expected for a non to low interacting system. In contrast, the absorption cross section of the fused dimer is changing significantly compared to monomers, on account of the wave-functions coupling and hybridization. The absorption cross section of fused dimers is losing the distinct features of the monomers. While at energies higher than 2.5eV$\sigma$, is approximately twice the value of the monomers. At the band edge, the $\sigma$ of the dimer is at the height of the monomer but broadened significantly towards the lower energies (red in Fig.5a). Integration on $\sigma$ of the band edge transition, leading to an estimation of the oscillator strength of the band gap transition, gives an overall value that is ~90% of the integrated cross section of the non-fused dimers. Similar results were reported for fusion of two PbSe Quantum dots[56].

To estimate the theoretical absorption spectrum, using Eq. 14 we have calculated the overlap integrals between the electron and hole states in the end of the Schrodinger-Poisson iteration for the band gap and excited states within this single valence band approximation. One can see that at higher energies the density of states for the dimer is much higher compared to the monomers, and considering the larger density of transitions which are partially allowed for dimers, one can understand the broadening leading to vanishing features in the fused dimers absorption spectrum (Fig.5b-c). At the band edge, the overlap integral of the dimer is slightly lower than the monomer but one should take into account the possibility for the photon to be absorbed in both cores. Hence, the integrated cross section of the dimer at the band edge should be slightly lower than twice the one of the monomer.

The broader absorption feature of fused dimers at the band edge can be understood as follows. As discussed before, the red shift of both the emission and the absorption is sensitive to the neck size. Small variations in the neck size among the fused dimers can lead to a significant inhomogeneous broadening of the band edge transition. In addition, unlike the emission, absorption is occurring also for higher transitions than the band edge. The light hole effective mass of CdSe is in the order of the electron suggesting that hybridization can also occur in the valence band states contributed by the light hole even at energies near the gap. In this case, one should expect two closely spaced allowed transitions for the bonding and anti-bonding levels of both electron and hole which can explain the broadening of the band-edge transitions. More elaborated calculation using valence band mixing as applied for colloidal quantum dots before can shed light on these transitions. This can be performed under the multiband 4 or 6 band $K \cdot P$ approximation that is beyond the scope of the single band approximation of this paper. These results suggest that the absorption spectrum is changing significantly upon fusion. This fact can be used to identify the fusion process and quantitative interpretation can be achieved with the more elaborate consideration of the band mixing effects.

**Conclusions**

In summary, we theoretically studied the electronic and optical signature of CQDM dimers with respect to their monomers building blocks. We examined the dependence of the coupling energies for different core diameters and barrier thicknesses assuming several conduction band offsets between CdSe to CdS. The results show that unlike the intuitive thinking that as the potential barrier height is lower the coupling energy will be higher, one should take into account the position of the energy eigen-value with respect to the potential barrier as related to the band offsets and effective mass of the specific semiconductor system. In some cases, a higher potential barrier can lead to higher coupling energies. Next, we showed that the neck in the interface between the two CQDs forming the molecule has a large impact on the coupling energy, and more so when the band offset is small. Then we showed that the fusion energy and the Coulomb energy between the electron to hole are acting in an opposite way in terms of the contribution to the optical band gap energy shift between monomers to dimers, and that the trend is changing non-monotonously as a function of the core diameter. Still, we predict that the fusion is always leading to red shifted emission and it is maximized for very small cores or a thin barrier thickness. Lastly, we show that the absorption cross section is changing upon fusion leading to a much larger absorption deep in the region where shell states also contribute compared to the band edge.

The main findings from the theoretical analysis with regards to the signatures of coupling are in agreement with the experimental observation for the CdSe/CdS dimer CQDMs. This includes the red shift of the band gap, the broadening of the transitions near the gap, and the significant change in the absorption spectrum at higher energies and at the band edge for the CQDMs. Such theoretical modelling is therefore an essential tool for the intelligent design of CQDMs constructed from the diverse selection of core/shell CQD monomer building blocks. It will also allow for tailoring of CQDMs for specific tasks relevant for applications in displays, sensing, fluorescence tagging and quantum technologies schemes.

### Supplementary material

See supplementary material for the monomer wave-functions after applying the Coulomb interaction and the method used to calculate the absorption cross section from the absorption spectrum and ICP-AES measurements. In addition, absorption cross section along with calculated transitions overlap integral for 0.32eV band offset is also provided.

### Acknowledgments

The research leading to these results has received financial support from the European Research Council (ERC) under the European Union's Horizon 2020 research and innovation programme (advanced investigator grant agreement No [741767]). U.B thanks the Alfred & Erica Larisch Memorial Chair. Y.E.P. acknowledges support by the Ministry of Science and Technology and the National Foundation for Applied and Engineering Sciences, Israel. J.B.C and S.K acknowledge the support from the Planning and Budgeting Committee of the higher board of education in Israel through a fellowship.

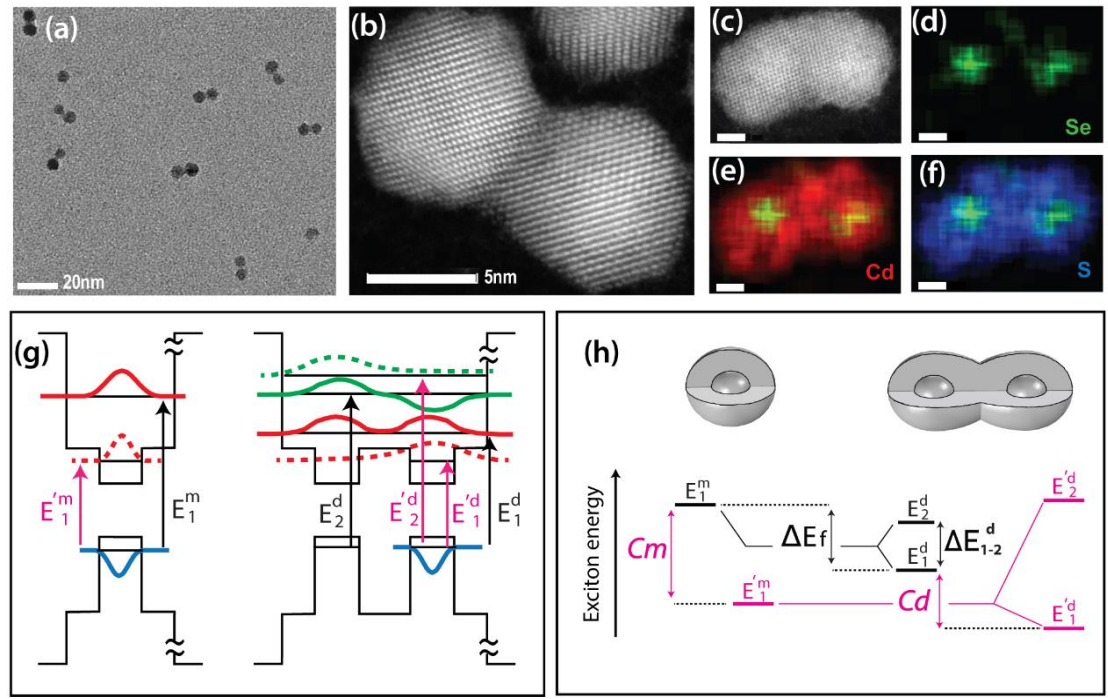

**Figure 1**. Coupled colloidal quantum dot molecules formed from coupling of two CQD CdSe/CdS core/shell monomers. (a) TEM image of CQDM comprised of two CdSe/CdS core/shell CQDs with 2.8nm core diameter and 2.1nm shell thickness. (b) HAADF-STEM image of a CQDM, (c)-(f) Energy dispersive X-Ray spectroscopy (EDS) mapping of the different elements. The CdSe cores are clearly resolved. Scale bars 2 nm. (g) scheme for the potential energy landscape of the monomer (left side) and the fused CQDM dimer (right side) along with a cross section of the highest hole wave-function (blue), the lowest state of the electron without Coulomb interaction (red), with coulomb (dashed red), the first excited state of the electron without Coulomb interaction (green) and with Coulomb (dashed green). (h) Exciton energy level ordering for the monomer CQD (left side) and the dimer CQDM (right side) before (black), and after (magenta) applying the coulomb interaction. $C_{m/d}$ refers to the coulomb energies in the monomer and the dimer respectively, $\Delta E_f$ refers to the fusion energy. $\Delta E_{1-2}^d$ refers to the coupling energy, the difference between the symmetric and anti-symmetric electron states in the dimer CQDM.

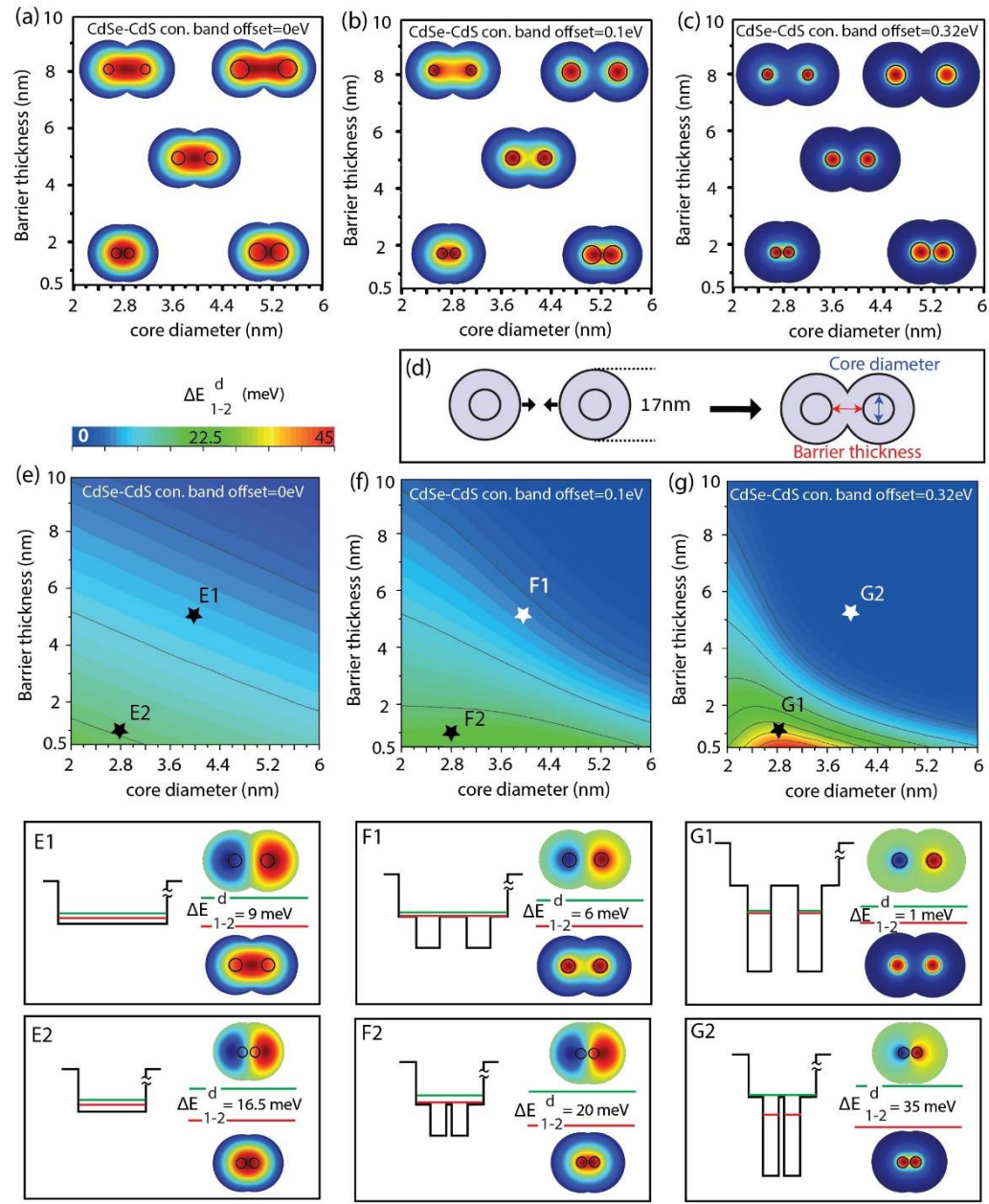

**Figure 2.** Effects of core dimensions, barrier thickness, and band offsets on the coupling of the lowest conduction band state in the CQDM. (a) The wave-functions of the first electronic state (symmetric state) of specific points on the contour graphs 2e-g. core diameter/barrier thickness 2.8nm/0.9nm, 2.8nm/8nm, 4nm/5nm, 5nm/0.9nm and 5nm/8nm in: (a) 0 eV conduction band offset, (b) 0.1 eV conduction band offset and (c) 0.32 eV conduction band offset. (d) The barrier width is controlled by overlapping the two outer spheres of the core/shell CQDs. A contour plot of the energy splitting between the symmetric and anti- symmetric states for the lowest conduction band levels in the CQDM (the energy values shown in the color scale) as a function of barrier thickness and core diameter in: (e) 0 eV, (f) 0.1 eV and (g) 0.32 eV conduction band offsets. E1, F1 and G1 refer to the point 4nm/5nm in graphs (e), (f) and (g) respectively, E2, F2 and G2 refer to the point 2.8nm/0.9nm in graphs (e), (f) and (g) respectively. The eigen-energies of the symmetric state (red line) and the anti-symmetric state (green line) of the electron with respect to the potential energy landscape together with their wave-functions is presented in the above points.

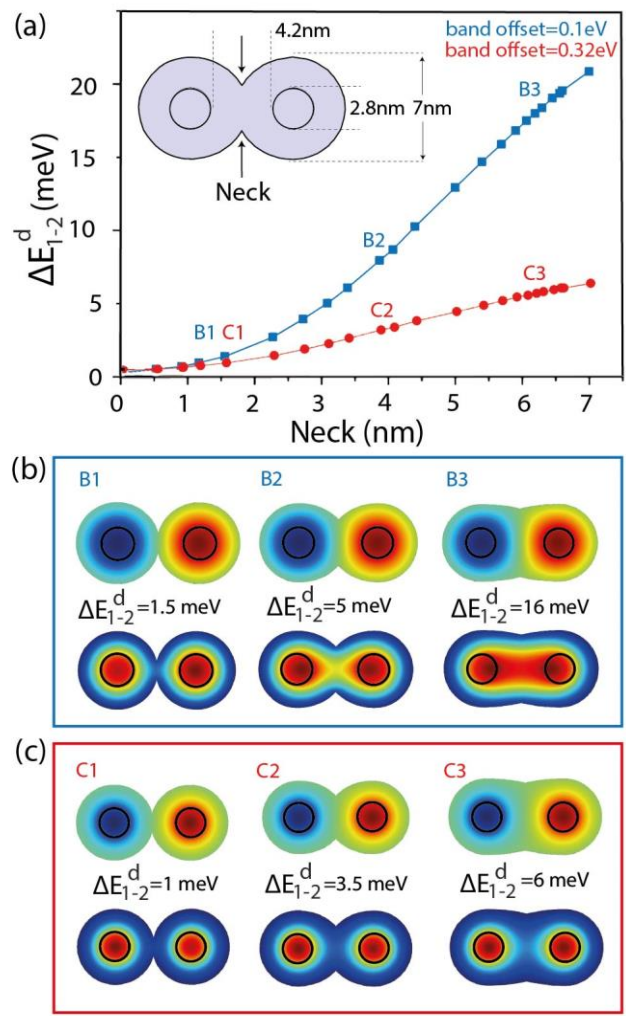

**Figure 3.** Neck size effect on the coupling of the lowest electron state in CQDMs. (a) Energy difference between the symmetric and anti- symmetric states as a function of the neck diameter. The blue curve refers to 0.1eV conduction band offset, and the red curve refers to 0.32eV conduction offset. Inset: the dimensions considered in the calculation. (b) The wave-functions of the symmetric (bottom) and the anti-symmetric (top) electronic states and the energy difference between them in three points: B1, B2 and B3 which refers to 1.5nm, 4nm and 6.2nm neck thickness, respectively (0.1nm band offset). (c) The wave-functions of the symmetric (bottom) and the anti- symmetric (top) electronic states and the energy difference between them in three points C1, C2 and C3 corresponding to 1.5nm, 4nm and 6.2nm neck thickness, respectively (0.32eV band offset).

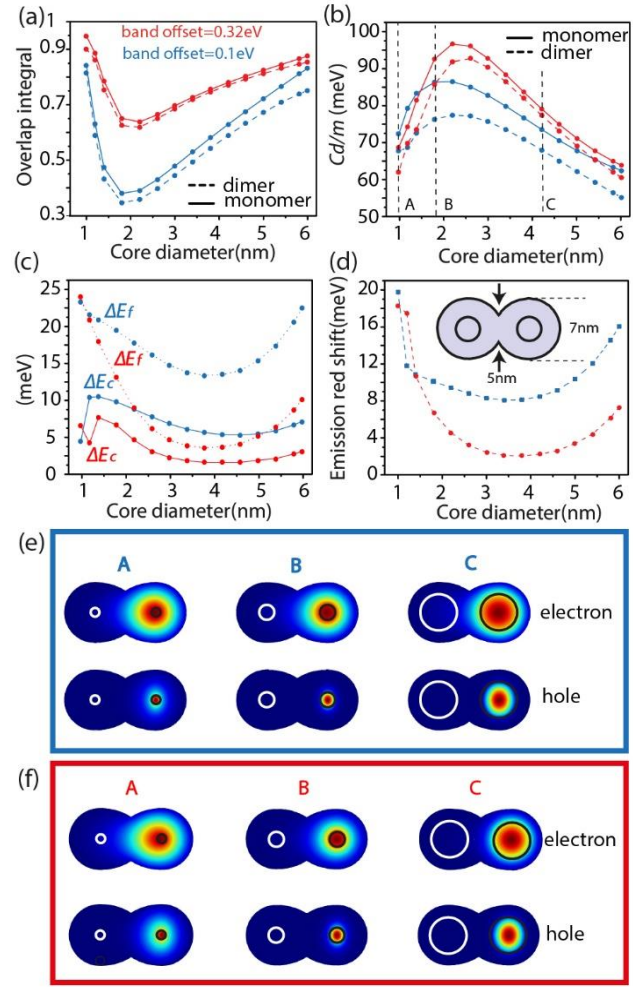

**Figure 4.** Excitonic behavior for the CQDM. (a) The overlap integral between the first electron and hole wave-function after applying the coulomb interaction as a function of the core diameter for the dimer (dashed line) and the monomer (solid line), considering 0.32eV (red) and 0.1eV (blue) conduction band offsets. (b) The coulomb energy $C_d$ of the dimer (dashed line) and monomer $C_m$ (solid line), in 0.32eV (red) and 0.1eV (blue) conduction band offsets as a function of the core diameter. (c) The difference between the monomer and dimer coulomb energy $\Delta E_c$ (solid line) and the fusion energy $\Delta E_f$ (dashed line) in 0.32eV (red) and 0.1eV (blue) conduction band offset as a function of the core diameter. (d) The emission red shift between the monomer and dimer in 0.32eV (red) and 0.1eV (blue) conduction band offset as a function of the core diameter. Inset: the dimensions considered in the calculation. (e-f) A, B and C show the wave-functions of the first electron and hole states after applying the coulomb interaction in 1nm, 1.8nm and 4.2nm core diameters, respectively, considering 0.1eV (blue frame) and 0.32eV (red frame) conduction band offsets.

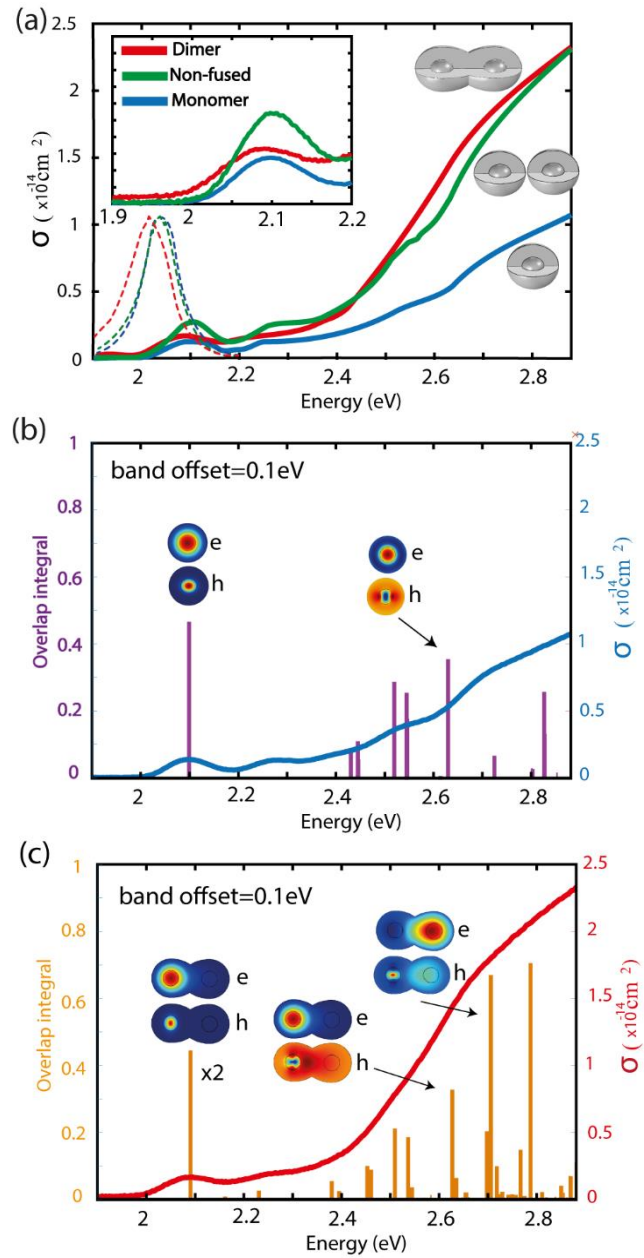

**Figure 5.** Absorption spectra of CQD versus CQDMs and comparing theory with experiments (for fused core diameter/shell thickness 2.8/2.1nm CdSe/CdS. (a) Absorption cross section as a function of energy for monomer CQD (blue), Non-fused CQD dimers (green) and fused dimer CQDM (red) together with the normalized emission spectrum (dashed line). (b) Monomer absorption cross section (blue) along with calculated transitions overlap integral (purple) for 0.1eV band offset. (c) Dimer absorption cross section (red) along with calculated transitions overlap integral (orange) for 0.1eV band offset. Electron and hole wave-functions involved in strong transitions are also presented.

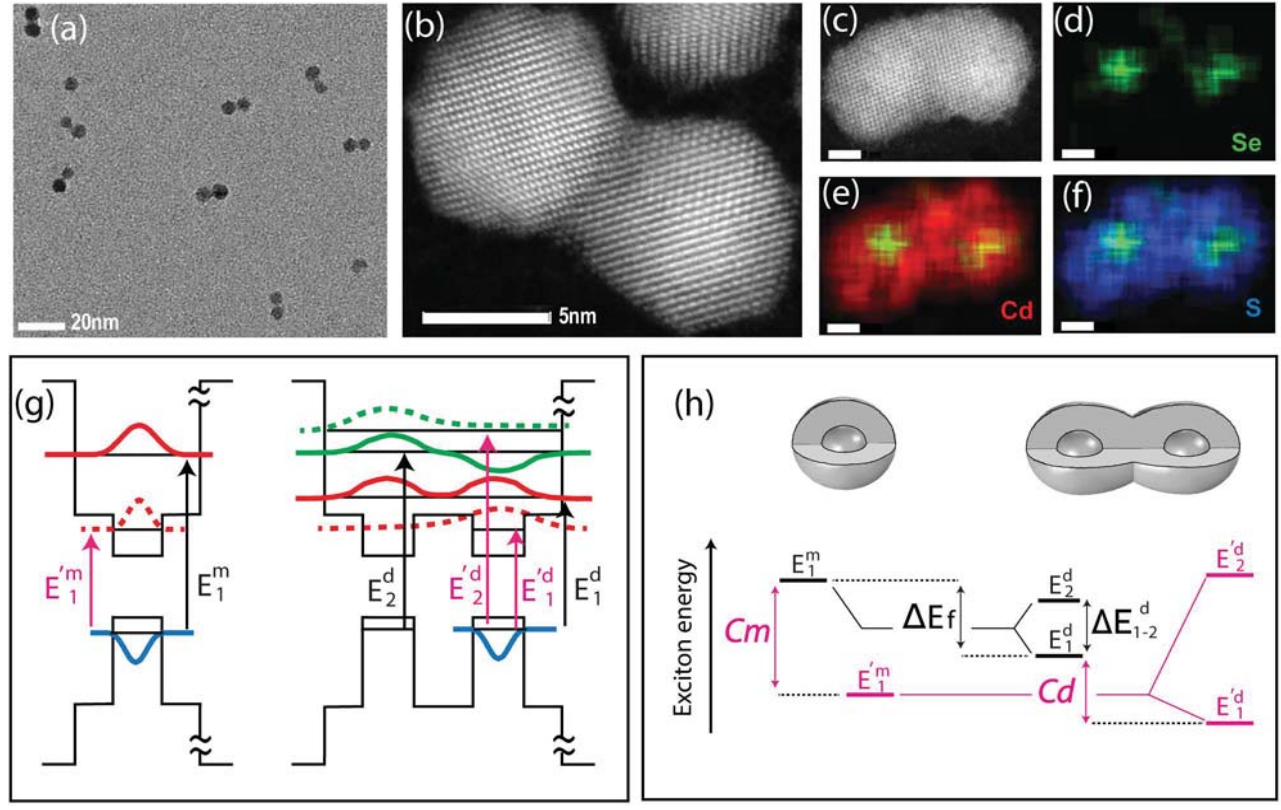

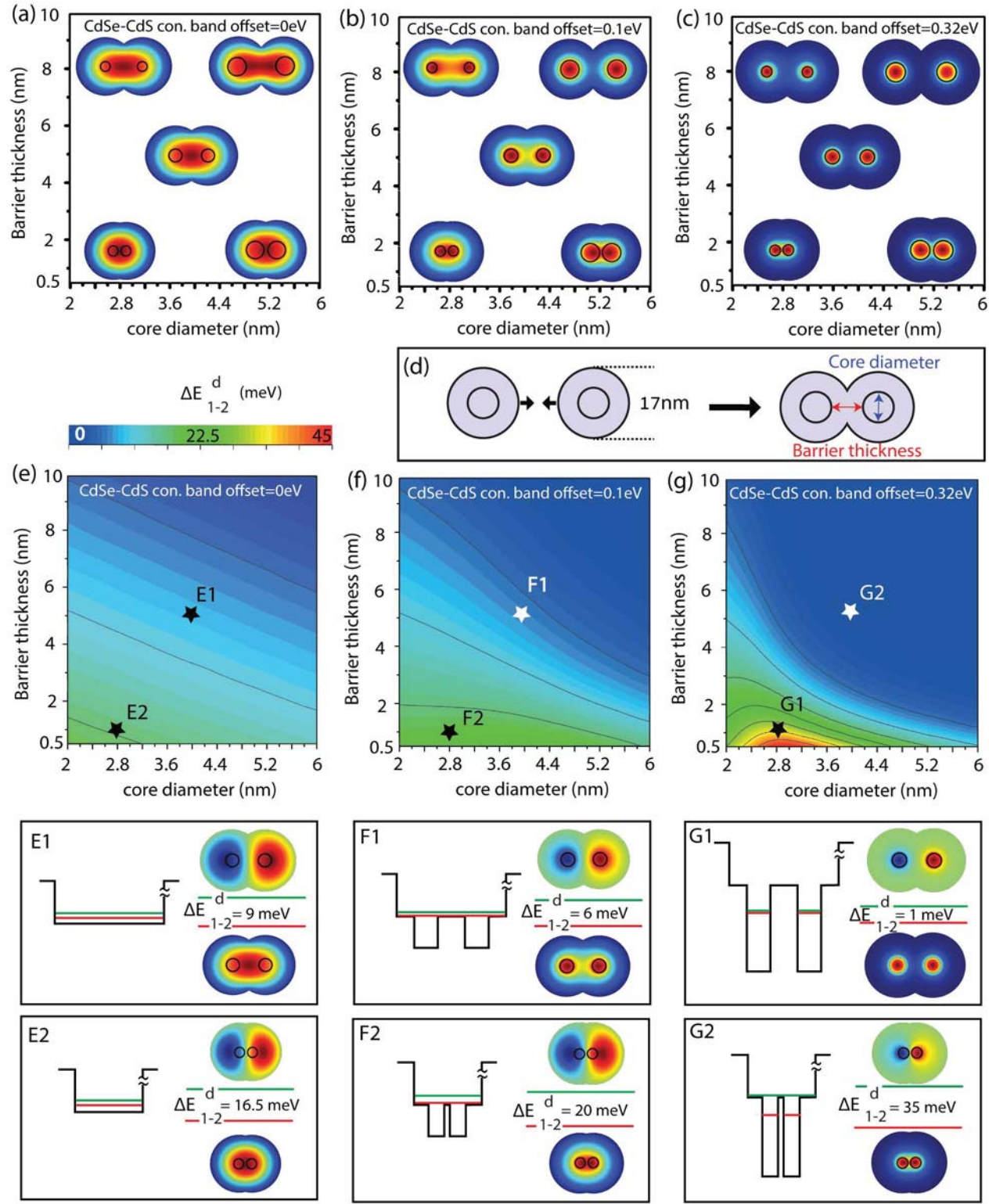

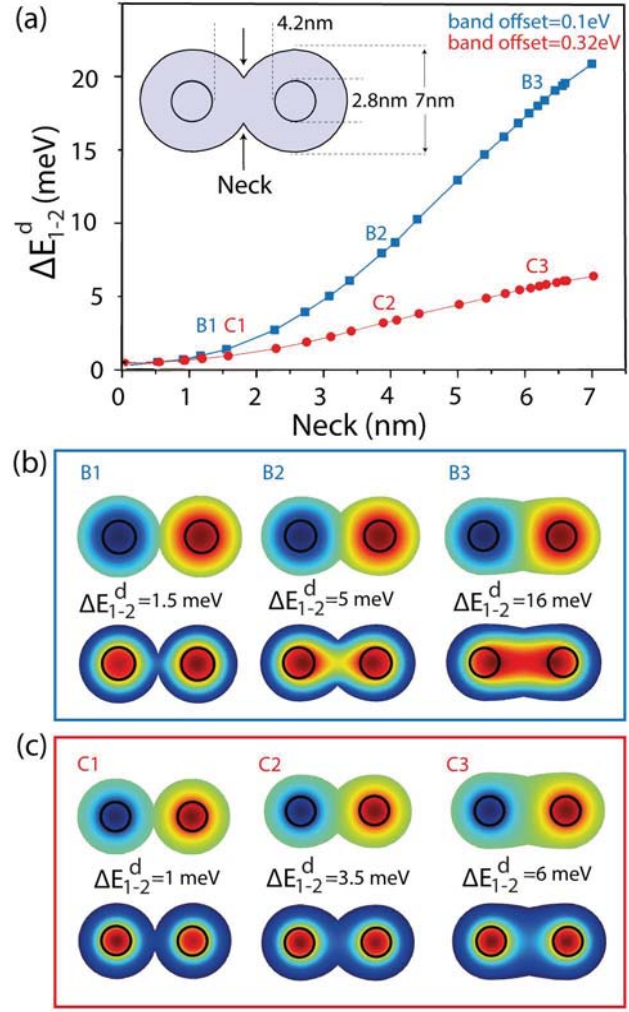

The Journal
of Chemical Physics

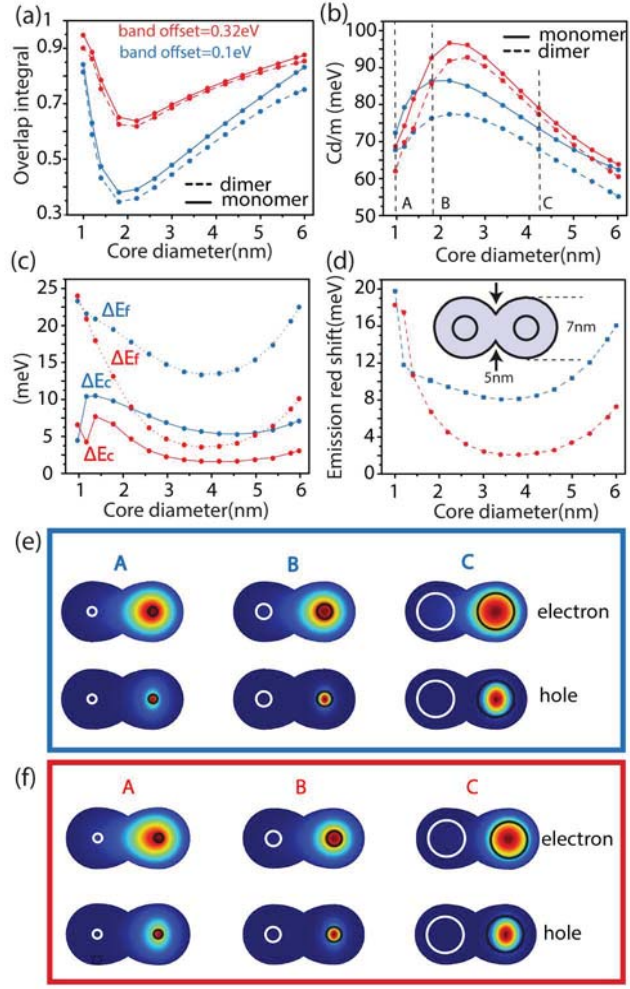

AIP
Publishing

The Journal
of Chemical Physics

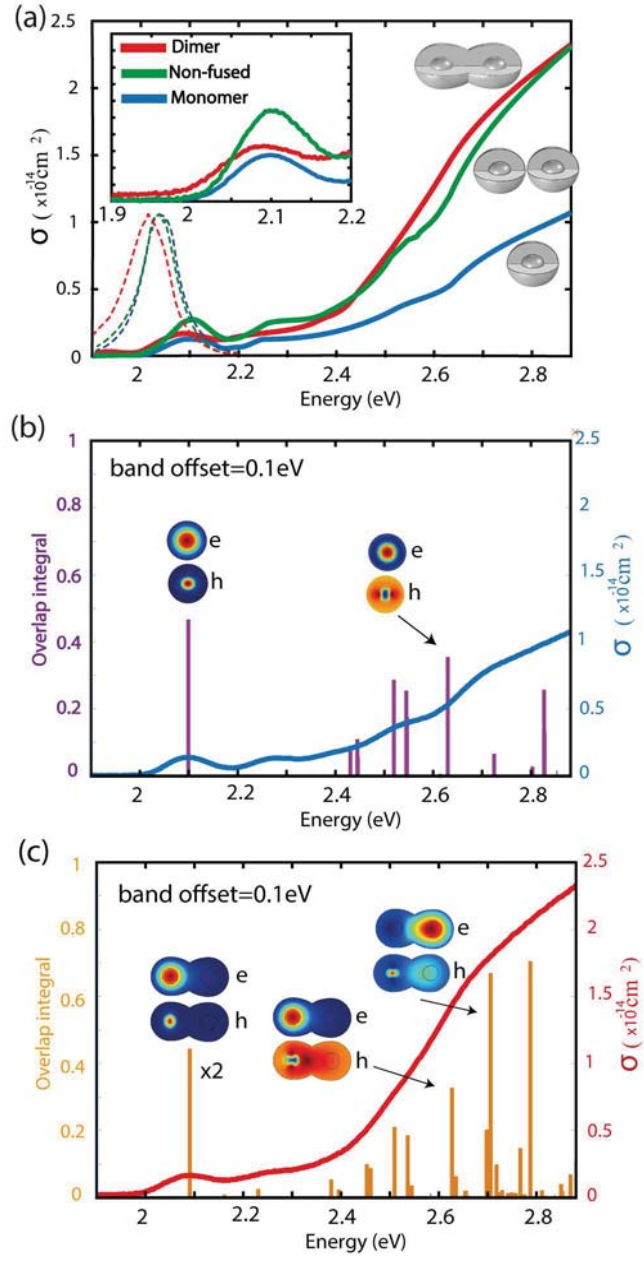